\documentclass[11pt]{article}

\usepackage[utf8]{inputenc}
\usepackage[T1]{fontenc}
\usepackage{lmodern}
\usepackage[margin=1in]{geometry}
\usepackage{microtype}
\usepackage{amsmath,amssymb,bm}
\usepackage{graphicx}
\usepackage[numbers,sort&compress,square]{natbib}
\usepackage[hidelinks]{hyperref}

\newcommand{\dd}{\mathrm{d}}
\newcommand{\eqdef}{\equiv}

\newcommand{\figfile}[2]{%
  \IfFileExists{#2.pdf}{\includegraphics[width=#1]{#2}}{%
  \IfFileExists{#2.png}{\includegraphics[width=#1]{#2}}{%
  \IfFileExists{#2.jpg}{\includegraphics[width=#1]{#2}}{%
  \IfFileExists{#2}{\includegraphics[width=#1]{#2}}{%
  \fbox{\parbox[c][0.25\textheight][c]{0.9\linewidth}{\centering\texttt{\detokenize{#2}}: figure file not found}}}}}}}

\title{A non-equilibrium thermodynamics formulation of shear set-points in adaptively remodeling vascular networks}
\author{Enrique Hernández-Lemus\\
\small Instituto Nacional de Medicina Genómica (INMEGEN), Periférico Sur 4809, Arenal Tepepan,\\
\small Tlalpan, 14610 Mexico City, Mexico\\
\small \texttt{ehernandez@inmegen.gob.mx}\\[0.5em]
Daniel Pérez-Calixto\\
\small Facultad de Ciencias, Universidad Nacional Autónoma de México, Ciudad Universitaria,\\
\small 04510 Mexico City, Mexico; Instituto Nacional de Medicina Genómica (INMEGEN), Mexico City, Mexico}
\date{}

\begin{document}
\maketitle

\begin{abstract}
Blood vessels remodel in response to wall shear stress while the network operates
  under a metabolic maintenance budget. For a power-law maintenance cost, minimizing
  dissipation at fixed budget yields a local shear set-point, and a purely local adaptation rule
  drives the network towards this optimum. Here we give this result a non-equilibrium thermodynamic
  formulation. Blood flow is an exactly linear Onsager system, and the slow remodeling of vessel radii
  is a gradient flow generated by a Rayleigh dissipation functional. This identifies the structural
  mobility, which scales inversely with the maintenance cost of each vessel and can be measured from
  radius fluctuations. Enforcing the budget exactly requires a time-dependent metabolic multiplier,
  which we obtain in closed form; with a fixed multiplier the budget is violated by up to one third
  during transients. The fluid displaced by remodeling vessels couples flow and structure through a
  symmetric, non-diagonal mobility, so that Onsager reciprocity becomes a testable statement:
  neighbouring vessels should show anticorrelated radius fluctuations. With noise, the structural
  sector obeys a second law and an integral fluctuation theorem. Non-reciprocal signaling between
  vessels, such as upstream conducted responses, then breaks the time-reversal
  symmetry of radius fluctuations and produces housekeeping entropy, detectable from time series of
  realistic length. Finally, we construct a partition function over network
  topologies with the number of loops as order parameter. Numerical tests show that loop counts do
  not collapse onto a single curve against an effective temperature, and that they
  depend strongly on the detection threshold.
\end{abstract}


\section{Introduction}
\label{sec:intro}

Vascular networks remodel locally (i.e., vessels dilate or constrict in response to wall shear stress), yet the efficiency of blood transport is a property of the network as a whole. Reconciling this
local sensing with a global energetic goal is a long-standing question in the physiology of
adaptive flow networks, and it applies equally to leaf venation and to engineered transport systems
\cite{davies1995,kamiya1980,pries1995,reneman2006}. A recent variational treatment resolved this
tension for a power-law maintenance cost: minimizing dissipation under a fixed maintenance budget
yields an exact local shear set-point that depends jointly on vessel radius and length, recovers
Murray's law as a special case, and is realized dynamically by a purely local adaptation rule that
performs descent on the associated global transport functional
\cite{murray1926,west1997,banavar2000,bohn2007,durand2007,vazquez2026}. Simulation of that rule
further showed that developmental history and fluctuating demand shape network topology: staged
growth with regrowth of pruned connections systematically outperforms adaptation from a dense
initial state, and demand fluctuations stabilize redundant loops without increasing the maintenance
budget \cite{corson2010,katifori2010,kaiser2020,martens2017}.\\

That analysis is thermodynamic in content but not in form: the local adaptation rule and its
associated Lyapunov inequality are asserted rather than derived from a variational dissipation
principle, no phenomenological (Onsager) coefficients are identified, and the topological
degeneracy and loop-selection phenomena are documented numerically without a statistical-mechanical
framework that would make them predictive beyond the simulated parameter range. The purpose of the
present paper is to supply that framework. We show that the vascular remodeling problem is a two-time-scale system whose fast sector is
exactly linear and whose slow sector is a gradient flow that reduces to linear response near the
set-point, in the sense of Onsager and Prigogine
\cite{onsager1931a,onsager1931b,prigogine1955,hu2013}, that its local adaptation rule can be written as a
gradient flow generated by a
Rayleigh dissipation functional together with the maintenance constraint, and that its
fluctuation-driven topological transitions admit a partition-function description formally
analogous to that of other adaptive network models
\cite{katifori2010,ronellenfitsch2016}.\\

The present work builds on several lines of research.
Flux-driven adaptation rules were introduced for \emph{Physarum} \cite{tero2007,tero2010} and for
microvascular networks, where the models of Pries, Secomb and coworkers already combine shear,
pressure, metabolic and conducted (non-local) stimuli \cite{pries1998,pries2001}; convergence of such
dynamics to optimal transport and the existence of Lyapunov functionals have been proved or
characterized in detail \cite{bonifaci2012,facca2018,facca2021}, and gradient flows that respect
constraints exactly have been constructed for microvascular optimization \cite{chang2019}. The shear
set-point itself has direct experimental support \cite{baeyens2015}, and set-points cued to shear
peaks have been shown to stabilize loopy capillary beds \cite{qi2024}. Time-dependent loads act
through an averaged dynamics with an explicit criterion for tree topologies \cite{lonardi2023};
topology selection under noise has been formulated in statistical-mechanical terms
\cite{palla2004,woodhouse2016,koehl2019}, and a curve collapse of fluctuation-induced topology
transitions has been reported for coupled adaptive networks \cite{kramer2020}. Global conservation of
fluid volume has been identified as the parameter that integrates network architecture into vein fate
in \emph{Physarum} \cite{marbach2023}. On the thermodynamic side, network thermodynamics provides the
classical Onsager structure on graphs \cite{oster1971,oster1973}, and local learning rules in physical
networks are a related instance of local rules descending global functionals \cite{stern2021}.
Finally, generalizations of Murray's law with non-cubic cost exponents
\cite{uylings1977,sherman1981,huo2012,marchesi2026} and empirical scalings of wall shear stress with
radius \cite{seymour2019} bear directly on the radius dependence of Eq.~\eqref{eq:setpoint}.\\

Gradient-flow representations of local adaptation rules have been discussed before
\cite{hu2013,ronellenfitsch2016,facca2018,facca2021}. What is new here is: (i) the structural
Onsager coefficient and its testable scaling with the maintenance cost; (ii) the exact enforcement of
the metabolic budget through a closed-form Gaussian multiplier; (iii) a hydraulic coupling between
vessels that makes Onsager reciprocity testable; and (iv) a second law for the structural sector,
which turns reciprocity into a time-reversal test for active signaling between vessels.\\

The paper is organized as follows. Section~\ref{sec:model} recalls the network model and the
central optimality result. Section~\ref{sec:fast} establishes the fast hydraulic sector as an exact
linear Onsager system. Section~\ref{sec:slow} writes the slow structural sector as a Rayleighian
gradient flow, obtains the structural Onsager coefficient in closed form, recovers the reported
Lyapunov inequality as a special case of the bilinear form $\sigma = \sum_e L_e X_e^2$, and derives
the hydraulic--structural coupling induced by the displaced volume. Section~\ref{sec:multiplier} examines the metabolic multiplier and
shows that its quenched treatment is a canonical-ensemble approximation to an exact
micro-canonical (Gaussian-constrained) dynamics, which we derive in closed form.
Section~\ref{sec:fluctuations} extends the theory to fluctuating demand via a Langevin equation and
constructs the associated partition function over network topologies, from which the loop count
emerges as a thermal order parameter; it also formulates the second law and the fluctuation theorem
of the structural sector and tests both predictions numerically. Section~\ref{sec:discussion} discusses the regime of validity
of the classical minimum-entropy-production theorem in this system, collects the falsifiable
predictions of the theory, and concludes. Appendix~\ref{app:derivations} collects the derivations step by
step, and Appendix~\ref{app:numerics} describes the numerical methods and parameters.

\section{Network model and the local optimality condition}
\label{sec:model}

We adopt the network model and notation of Ref.~\cite{vazquez2026}. Consider a connected graph with
$N$ sink nodes, a single source, and $n = N+1$ nodes in total; each sink withdraws one unit of flow.
An edge $e$ has radius $r_e$, length $\ell_e$, Poiseuille conductance $w_e = r_e^4/\ell_e$ (in
reduced units absorbing viscosity and geometric prefactors), signed flow $f_e$, and wall shear
stress $\tau_e = f_e/r_e^3$. Writing $b_e$ for the oriented incidence vector of edge $e$ and
$L = \sum_e w_e\, b_e b_e^{\mathsf T}$ for the weighted graph Laplacian, node pressures are
$\phi = L^{+} I$, where $I$ is the injection vector and $L^{+}$ the Moore--Penrose pseudoinverse.
The total dissipated power is
\begin{equation}
  D = \sum_e \frac{f_e^2}{w_e},
  \label{eq:D}
\end{equation}
and the maintenance cost, for vessel mass $m_e \propto r_e^2 \ell_e$ raised to a power $b$, is
\begin{equation}
  C_b = \sum_e c_e w_e^{\alpha}, \qquad \alpha = \frac{b}{2}, \qquad c_e = \ell_e^{3b/2},
  \qquad 0 < b < 2,
  \label{eq:Cb}
\end{equation}
and we denote by $C_e \eqdef c_e w_e^{\alpha} = r_e^{2b}\ell_e^{\,b}$ the maintenance cost of
edge $e$, so that $C_b=\sum_e C_e$. The budget is fixed, $C_b = C_0$. Minimizing $D$ at fixed $C_b$ by a Lagrange
multiplier $\lambda$ yields, on the active support of a tree, the shear set-point
\begin{equation}
  |\tau_e| = \tau_{\rm set}(r_e,\ell_e) = \tau_0\, r_e^{\,b-1}\ell_e^{(b-1)/2}, \qquad
  \tau_0 = \sqrt{\lambda \alpha},
  \label{eq:setpoint}
\end{equation}
which reduces to Murray's uniform-shear condition at $b=1$. Equivalently, the stationarity
condition $\partial D/\partial w_e + \lambda\,\partial C_b/\partial w_e = 0$ reads
$(\Delta p_e)^2 = \lambda\alpha\, C_e/w_e$. Writing $x_e \eqdef \ln r_e$ and the
normalized shear $z_e \eqdef |\tau_e|/\tau_{\rm set} \ge 0$, the reported local adaptation rule is
\begin{equation}
  \dot x_e = \kappa\,(z_e - 1), \qquad \kappa > 0,
  \label{eq:rule}
\end{equation}
under which the composite functional
\begin{equation}
  \Phi(x) \eqdef D(x) + \lambda\, C_b(x)
  \label{eq:Phi}
\end{equation}
obeys
\begin{equation}
  {\dot\Phi = -2b\lambda\kappa \sum_e C_e\,(z_e-1)^2 (z_e+1) \le 0.}
  \label{eq:Phidot_reported}
\end{equation}

Equations~\eqref{eq:rule}--\eqref{eq:Phidot_reported} are the starting point of the present analysis:
we now show that they are not independent postulates but admit an Onsager--Rayleigh gradient-flow representation once the network is separated into its fast and slow sectors.

\section{The fast hydraulic sector as an exact linear Onsager system}
\label{sec:fast}

For fixed geometry $\{w_e\}$, Poiseuille's law $f_e = w_e\, b_e^{\mathsf T}\phi$ is already a linear
force--flux relation, with flux $f_e$, thermodynamic force $\Delta p_e = b_e^{\mathsf T}\phi$ (the
pressure drop across edge $e$), and phenomenological coefficient $w_e > 0$. The corresponding
entropy-production (dissipation) rate is the bilinear Onsager form
\begin{equation}
  \sigma_{\rm fast} = \sum_e f_e\, \Delta p_e = \sum_e \frac{f_e^2}{w_e} = D \ge 0.
  \label{eq:sigma_fast}
\end{equation}
The dissipation rate is thus the total power of Eq.~\eqref{eq:D}. (Strictly, the entropy production is $D/T$, with $T$ the temperature of the fluid; in reduced
units $T$ is absorbed and we identify both.)
Because $L = \sum_e w_e\, b_e b_e^{\mathsf T}$ is a sum of symmetric outer products, the Onsager
reciprocity relations for this sector are satisfied identically, with no additional symmetry
assumption required. Moreover, for fixed $\{w_e\}$ and fixed sink demands, $D$ is a strictly convex
functional of the flow field subject to Kirchhoff's current law, so the physically realized flow is
exactly its unique minimizer: this is Thomson's principle, the linear-network counterpart of the
minimum-entropy-production theorem of Prigogine \cite{prigogine1955}. A useful corollary, used in Section~\ref{sec:slow}, follows from the envelope
theorem: since the flow minimizes $D$ at fixed demands, $\partial D/\partial w_e = -(\Delta p_e)^2$,
with no contribution from the implicit dependence of the flows on $w_e$. The reduced dissipation $D(x)$ appearing in $\Phi$, Eq.~\eqref{eq:Phi}, is
therefore the exact result of adiabatically eliminating the fast sector: introducing
$\varepsilon = \tau_{\rm hyd}/\tau_{\rm remod} \ll 1$ and writing $\varepsilon\,\dot\phi = I - L\phi = -\partial\mathcal{J}/\partial\phi$,
with $\mathcal{J}(\phi) = \tfrac12\phi^{\mathsf T}L\phi - I^{\mathsf T}\phi$ the hydraulic co-content,
the limit $\varepsilon \to 0$ collapses to the algebraic constraint
$\phi = L^{+}I$ used throughout Section~\ref{sec:model}. The physiological separation of time scales
between flow relaxation (seconds) and structural remodeling (hours to days) justifies this
elimination a priori.

\section{The slow structural sector: a Rayleighian formulation of the adaptation rule}
\label{sec:slow}

\subsection{The Rayleigh dissipation functional}

We now show that Eq.~\eqref{eq:rule} has the structure of a gradient flow generated by an Onsager variational
principle \cite{onsager1931a,onsager1931b,doi2011,doi2013}. Define the Rayleighian \cite{Beris2024}
\begin{equation}
  \mathcal{R}(\dot x; x) = \sum_e \frac{\partial \Phi}{\partial x_e}\,\dot x_e
  \;+\; \frac12 \sum_e \frac{\dot x_e^{\,2}}{L_e(x)},
  \label{eq:rayleighian}
\end{equation}
in which the first term is the instantaneous rate of change of the free-energy-like functional
$\Phi$ and the second is a quadratic dissipation function penalizing the rate of structural change,
with $L_e(x) > 0$ an as-yet-unspecified structural mobility. The physical trajectory extremizes
$\mathcal{R}$ over $\dot x$ at fixed $x$; stationarity of Eq.~\eqref{eq:rayleighian}, $\partial \mathcal{R}/\partial \dot x_e = 0$,
gives the general force--flux law
\begin{equation}
  \dot x_e = -L_e\, \frac{\partial \Phi}{\partial x_e} \eqdef L_e X_e,
  \qquad X_e \eqdef -\frac{\partial \Phi}{\partial x_e}.
  \label{eq:onsager_slow}
\end{equation}

\subsection{The structural Onsager coefficient}

Because $w_e = r_e^4/\ell_e$ at fixed $\ell_e$, $\dd w_e/\dd x_e = 4 w_e$. Using
$\partial D/\partial w_e = -(\Delta p_e)^2$ and the set-point relation
$(\Delta p_e)^2 = z_e^2\, \lambda \alpha\, C_e/w_e = z_e^2\,\lambda\alpha\, c_e w_e^{\alpha-1}$ (from Eqs.~\eqref{eq:Cb}--\eqref{eq:setpoint}),
\begin{equation}
  \frac{\partial \Phi}{\partial x_e} = 4 w_e\, \frac{\partial \Phi}{\partial w_e}
  = {4\lambda\alpha\, C_e\,(1 - z_e^2) = -2\lambda b\, C_e\,(z_e - 1)(z_e+1),}
  \label{eq:dPhidx}
\end{equation}
so that the thermodynamic force conjugate to $x_e$ is
\begin{equation}
  X_e = 2\lambda b\, {C_e}\,(z_e-1)(z_e+1).
  \label{eq:force}
\end{equation}
Matching Eq.~\eqref{eq:onsager_slow}, with the force of Eq.~\eqref{eq:force}, to the reported rule, Eq.~\eqref{eq:rule}, fixes the structural
Onsager coefficient uniquely:
\begin{equation}
  L_e(x) = \dfrac{\kappa}{2\lambda b\, {C_e}\,(z_e+1)}{.}
  \label{eq:Le}
\end{equation}
Two remarks qualify this result. First, the argument runs from the rule to the mobility: given
Eq.~\eqref{eq:rule}, $L_e$ is fixed uniquely, but conversely any rule
$\dot x_e = \kappa F(z_e)$ with $\operatorname{sign}F = \operatorname{sign}(z_e-1)$ admits a
representation of the form~\eqref{eq:onsager_slow} with a positive mobility. The physical content of
Eq.~\eqref{eq:Le} is therefore not the rule itself, but the identification of its mobility, which
carries testable information. Second, $L_e$ depends on $z_e$, and hence on $X_e$ itself, so the
structural force--flux relation is nonlinear; it reduces to linear Onsager response, with the
state-independent coefficient $L_e^{(0)} = \kappa/(4\lambda b\, C_e)$, only in the neighbourhood of
the set-point $z_e = 1$.\\

Three consequences follow immediately. First, $L_e>0$ for all admissible states, since $\kappa$,
$\lambda$, $b$, $C_e>0$ and $z_e \ge 0$; this positivity is inherited from the positivity of the
maintenance cost and the multiplier, not imposed separately, and it is what guarantees the Lyapunov property below. Second, $L_e \propto C_e^{-1} = r_e^{-2b}\ell_e^{-b}$: vessels with larger local maintenance cost are
structurally stiffer, i.e.\ their radius responds more slowly per unit thermodynamic force. This is
a falsifiable prediction not present in the original formulation: fitted structural mobilities
should scale inversely with local maintenance cost across a vascular bed.\\

Third, $\lambda$ enters
$L_e$ exactly as an intensive potential conjugate to the extensive budget $C_0$, structurally
analogous to a chemical potential in a grand-canonical ensemble.\\

Figure~\ref{fig:mobility}A shows how the second consequence can be tested. The
fluctuation--dissipation relation of Section~\ref{sec:fluctuations} links the mobility to the
short-time diffusivity of the logarithmic radius, $\mathcal D_e = \langle(\Delta x_e)^2\rangle/(2\Delta t)
= T_{\rm eff}L_e$. The prediction $\mathcal D_e \propto C_e^{-1}$ can therefore be checked from radius
time series, without knowing $\lambda$ or $X_e$. In Langevin simulations around the optimal tree of a
$6\times 6$ jittered lattice, the fitted exponents are $-0.998$, $-1.003$ and $-0.997$ for $b = 0.5$, $1$
and $1.5$. A fit against $r_e$ alone gives a biased exponent ($-2.86$ instead of $-3$ at $b=1.5$),
because $C_e$ also depends on the vessel length, which must therefore be included. In simulation this
is a consistency check, since the rule defines $L_e$. Its empirical content is that $\kappa$ is the same
for all vessels, so that intravital measurements of radius fluctuations across a vascular bed should
give a log--log slope of $-1$ against $C_e$, and of $-2b$ against $r_e$ at fixed $\ell_e$.

\subsection{Dissipation function and Lyapunov stability}

Combining Eqs.~\eqref{eq:onsager_slow}--\eqref{eq:Le},
\begin{equation}
  \sigma_{\rm slow} \eqdef -\dot\Phi = \sum_e X_e \dot x_e = \sum_e L_e X_e^2
  = 2b\lambda\kappa \sum_e {C_e}\,(z_e-1)^2(z_e+1) \;\ge 0,
  \label{eq:sigma_slow}
\end{equation}
which reproduces Eq.~\eqref{eq:Phidot_reported} term for term. In the present derivation, however,
non-negativity follows directly from the bilinear Onsager form $\sigma = \sum_e L_e X_e^2$ together
with $L_e>0$, rather than from a separate convexity argument. Equality in Eq.~\eqref{eq:sigma_slow}
holds only at $z_e = 1$ for every active edge, i.e.\ the shear set-point of Eq.~\eqref{eq:setpoint}
is exactly the zero-dissipation manifold of the slow sector.
We stress that $\sigma_{\rm slow}$ has the bilinear Onsager form but is not a thermodynamic
entropy production in the strict sense: $\Phi$ is itself a rate of dissipation, so $-\dot\Phi$ has
units of power per unit time. Its non-negativity is the structural-sector analogue of the second law,
namely a Lyapunov property of the minimum-dissipation principle, in the spirit of Prigogine's
theorem and of the Glansdorff--Prigogine evolution criterion \cite{glansdorff1971}. A dimensionally consistent second law, in which $\Phi$ is converted into an energy and the
configurational entropy of the structural state is included, is formulated in
Section~\ref{sec:secondlaw}.

\subsection{Reciprocity}
\label{sec:reciprocity}

The structural Onsager matrix implied by Eq.~\eqref{eq:onsager_slow} is diagonal,
$L_{ee'} = L_e\,\delta_{ee'}$, reflecting the assumption of purely local stress sensing; Onsager
reciprocity, $L_{ee'} = L_{e'e}$, then holds trivially. A genuine test of this assumption would
require detecting cross-couplings $L_{ee'}$ with $e
e e'$ (i.e. remodeling at one vessel driven by
shear elsewhere via long-range mechanotransductive signaling) which the present rule excludes by
construction and which we leave for future work as a natural extension of the theory. Part of the
cross-couplings, however, is not an additional assumption: it follows from hydraulics alone, as shown
next.

\subsection{Hydraulic--structural coupling through the displaced volume}
\label{sec:coupling}

The separation into sectors of Sections~\ref{sec:fast}--\ref{sec:slow} neglects the fact that a
remodeling vessel changes its lumen volume, $v_e = \nu\, r_e^2\ell_e$ (with $\nu$ absorbing geometric
prefactors), and must therefore exchange fluid with the network. Mass conservation at the nodes reads
$(Bf)_i + (A\dot v)_i = I_i$, where $A_{ie} = \tfrac12|B_{ie}|$ assigns half of the volume change of each
vessel to each of its end nodes. Treating the node pressures as Lagrange multipliers of this
constraint, the joint Rayleighian of both sectors is
\begin{equation}
  \mathcal R(f,\dot x;\phi) = \frac12\sum_e\frac{f_e^2}{w_e} + \frac12\sum_e\frac{\dot x_e^{\,2}}{\Gamma_e}
  + \sum_e\frac{\partial\tilde\Phi}{\partial x_e}\,\dot x_e
  - \sum_i\phi_i\Big[(Bf)_i + (A\dot v)_i - I_i\Big],
  \label{eq:R_joint}
\end{equation}
where $\tilde\Phi = \tau_\Phi\Phi$ is the structural potential expressed as an energy, with $\tau_\Phi$ a
metabolic time scale converting the power $\Phi$ into an energy, and $\Gamma_e$ is the bare structural
mobility. The dissipation $D$ thus appears twice: as the physical dissipation of the hydraulic sector,
first term, and, evaluated on the static flow, as part of the biological target $\tilde\Phi$ that drives
remodeling; the second role is phenomenological and is what makes the structural drive active. Stationarity with respect to $f_e$ returns Poiseuille's law, and stationarity with respect to
$\dot x_e$ gives
\begin{equation}
  \dot x_e = \Gamma_e\big(X_e + v'_e\,\bar p_e\big), \qquad
  v'_e \eqdef \frac{\partial v_e}{\partial x_e} = 2v_e, \qquad \bar p_e \eqdef (A^{\mathsf T}\phi)_e,
  \label{eq:wall}
\end{equation}
with $X_e = -\partial\tilde\Phi/\partial x_e$. The same coefficient $v'_e$ that turns wall motion into a
volume source in the continuity equation turns the mean luminal pressure into a force on the wall:
this is Onsager reciprocity between the two sectors, obtained from the variational structure rather
than imposed. We assume that the static transmural load is balanced by wall tension (Laplace's law) and is absorbed
into the reference state, so only the pressure generated by the wall motion itself remains,
$L_{\rm r}\phi = -A_{\rm r}V'\dot x$, where the subscript r denotes removal of the source node, whose
pressure is fixed, and $V' = \operatorname{diag}(v'_e)$. Eliminating it,
\begin{equation}
  \big(\Gamma^{-1} + K\big)\,\dot x = X, \qquad
  K \eqdef V' A_{\rm r}^{\mathsf T} L_{\rm r}^{-1} A_{\rm r} V',
  \label{eq:coupled}
\end{equation}
so that the effective structural Onsager matrix is $L^{\rm eff} = (\Gamma^{-1}+K)^{-1}$.
The assumption is not innocuous: the static mean pressure depends on the whole configuration $x$, and
the force $v'_e\bar p^{(0)}_e(x)$ is in general not a gradient, so that, if it is not fully balanced by
the wall, it adds a non-conservative drive that breaks detailed balance (Section~\ref{sec:secondlaw}).\\

\begin{figure}[t]
  \centering
  \figfile{\textwidth}{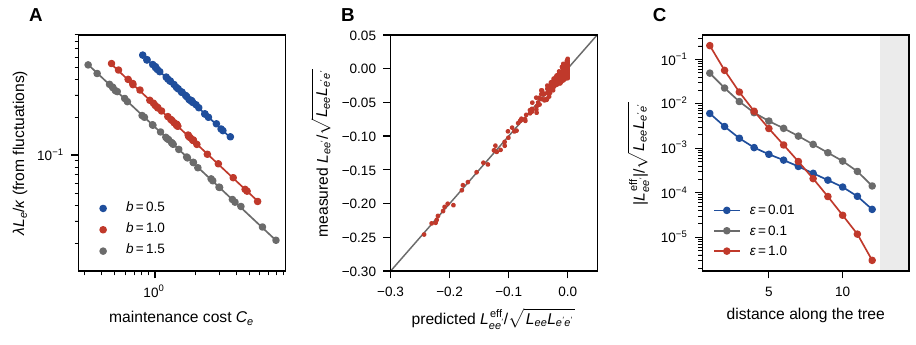}
  \caption{Structural mobilities obtained from fluctuations, in Langevin simulations around the
  optimal tree of a $6\times 6$ jittered lattice with the source at a corner. (A) Short-time
  diffusivity of $\ln r_e$ divided by $T_{\rm eff}$, which equals the mobility $L_e$, against the
  maintenance cost $C_e$ of each vessel, without hydraulic coupling. Lines: Eq.~\eqref{eq:Le} at
  $z_e=1$. (B) Cross-mobilities measured from the covariance of radius increments against the
  prediction $L^{\rm eff}=(\Gamma^{-1}+K)^{-1}$, Eq.~\eqref{eq:coupled}, for $b=1$ and
  $\varepsilon=0.3$; both are normalized by $\sqrt{L_{ee}L_{e'e'}}$, and the line marks equality.
  (C) Normalized cross-mobility against the distance between vessels along the tree, for three
  coupling strengths. The shaded band contains pairs in branches that meet only at the source, where
  the coupling vanishes.}
  \label{fig:mobility}
\end{figure}

Four properties follow. (i) $L^{\rm eff}$ is symmetric and positive definite, because $L_{\rm r}^{-1}$
is; reciprocity, $L^{\rm eff}_{ee'} = L^{\rm eff}_{e'e}$, now holds for a non-diagonal matrix and is
therefore a non-trivial, testable statement. (ii) To first order in the coupling, $L^{\rm eff}_{ee'}\simeq-\Gamma_eK_{ee'}\Gamma_{e'}<0$ for every pair
of vessels in the same branch, because $L_{\rm r}^{-1}$ has positive entries on each connected
component (Appendix~\ref{app:sign}). Physically, a dilating vessel draws fluid from its
neighbourhood, lowers the local pressure and drives the contraction of hydraulically connected
vessels. Numerically, the sign persists up to $\varepsilon\approx1$ in all 72 cases tested
(12 networks, 3 values of $b$) and breaks down only at strong coupling, $\varepsilon = 10$. The
coupling decays with distance along the tree and vanishes between branches that meet only at the
pressure-fixed source (Figure~\ref{fig:mobility}C). (iii) The coupling strength is set by $\varepsilon_e \eqdef \Gamma_e
K_{ee}$, the ratio between the hydraulic drag time of wall motion and the remodeling time, of the same order as the small parameter $\varepsilon$ of Section~\ref{sec:fast}. To first order,
$L^{\rm eff} \simeq \Gamma - \Gamma K\Gamma$ is the leading correction to the adiabatic elimination:
it is small in mature vasculature (seconds against hours to days), but not in systems in which flow and
adaptation evolve on comparable time scales, such as \emph{Physarum} or rapidly growing plexuses.
(iv) The uncoupled theory only fixes the product $\tau_\Phi\Gamma_e = L_e$ of Eq.~\eqref{eq:Le}; the
coupling makes $\tau_\Phi$ and $\Gamma_e$ separately identifiable.\\

Multiplying Eq.~\eqref{eq:coupled} by $\dot x$ gives the power balance of the coupled system at fixed
$\lambda$ (with the Gaussian multiplier of Section~\ref{sec:multiplier}, $\tilde\Phi$ is replaced by
$\tau_\Phi D$),
\begin{equation}
  -\frac{\dd\tilde\Phi}{\dd t} = \sum_e\frac{\dot x_e^{\,2}}{\Gamma_e} + \dot x^{\mathsf T}K\dot x
  = \sum_e\frac{\dot x_e^{\,2}}{\Gamma_e} + \sum_e\frac{(f_e^{\rm dyn})^2}{w_e} \;\ge 0,
  \label{eq:balance_coupled}
\end{equation}
where $f^{\rm dyn}$ are the flows induced by wall motion. Part of the decrease of the structural
potential is therefore dissipated viscously in the lumen, and the entropy production of the slow sector
has a genuinely hydraulic component. Along relaxations towards the optimal tree the balance is
satisfied numerically to $0.2\%$ (the finite-difference error), with a hydraulic share of $0.8\%$, $20\%$
and $53\%$ for $\varepsilon = 0.01$, $0.1$ and $1$. The cross-mobilities can be measured from the
covariance of radius increments, $\langle\Delta x_e\,\Delta x_{e'}\rangle = 2T_{\rm eff}L^{\rm eff}_{ee'}\Delta
t$ (negative for neighbouring vessels, whose increments are therefore anticorrelated). In simulation,
the measured and predicted values agree with a correlation of $0.993$ (Figure~\ref{fig:mobility}B).

\section{The metabolic multiplier: quenched versus dynamically constrained dynamics}
\label{sec:multiplier}

Equation~\eqref{eq:sigma_slow} treats $\lambda$ as a constant. Differentiating the budget,
\begin{equation}
  \dot C_b = \sum_e \frac{\partial C_b}{\partial x_e}\dot x_e
  = 4\alpha\kappa \sum_e {C_e}\,(z_e - 1),
  \label{eq:Cbdot}
\end{equation}
which vanishes only at the fixed point, not along the trajectory: a constant $\lambda$ enforces the
budget constraint only asymptotically, and only if $\lambda$ is set to its optimal value
$\lambda^\star(C_0)$; for any other value the dynamics converges to the optimum of a different budget in a canonical-ensemble sense, analogous to running
constrained dynamics with a frozen Lagrange multiplier rather than a holonomic constraint force
solved self-consistently at every instant (as in the SHAKE/RATTLE algorithms or a Gaussian
isokinetic thermostat).

We now derive the exact constrained dynamics. Define the $\lambda$-independent gradients
\begin{equation}
  g_e \eqdef \frac{\partial D}{\partial x_e} = {-4 w_e (\Delta p_e)^2}, \qquad
  h_e \eqdef \frac{\partial C_b}{\partial x_e} = 4\alpha\, {C_e},
  \label{eq:gh}
\end{equation}
so that $\partial\Phi/\partial x_e = g_e + \lambda(t) h_e$ and the dynamics reads
$\dot x_e = -L_e\left[g_e + \lambda(t) h_e\right]$,
with $L_e$ given by Eq.~\eqref{eq:Le} evaluated at $\lambda(t)$.
Introducing, with the gradients of Eq.~\eqref{eq:gh}, the local, $\lambda$-independent shear ratio
$s_e \eqdef (-g_e/h_e)^{1/2} = [\,w_e(\Delta p_e)^2/(\alpha C_e)\,]^{1/2}$, the normalized shear is
$z_e = s_e/\sqrt{\lambda}$, and the rule~\eqref{eq:rule} reads $\dot x_e = \kappa(s_e/\sqrt\lambda - 1)$.
Requiring that Eq.~\eqref{eq:Cbdot} vanish at all times, $\dot C_b = \sum_e h_e \dot x_e \equiv 0$, then gives the multiplier in explicit form,
\begin{equation}
  {\sqrt{\lambda(t)} = \frac{\sum_e C_e\, s_e}{\sum_e C_e},}
  \label{eq:lambda_t}
\end{equation}
which is the Gaussian-constraint multiplier for this network: $\sqrt{\lambda(t)}$ is the
cost-weighted mean of the local shear ratios, determined instantaneously by the current flow field. Substituting back,
\begin{equation}
  {\dot x_e = \kappa\left(\frac{s_e\sum_{e'}C_{e'}}{\sum_{e'}C_{e'}s_{e'}} - 1\right),}
  \label{eq:projected_flow}
\end{equation}
which is a projected gradient flow: the multiplier is chosen at every instant so that the velocity is tangent to the constraint
manifold, $h^{\mathsf T}\dot x = 0$, i.e.\ orthogonal in the metric $L^{-1}$ to the Riemannian gradient
$Lh$ of the constraint, exactly as in constrained steepest descent \cite{chang2019}. Because
$\sum_e h_e\dot x_e = 0$, the time dependence of $\lambda$ does not enter the rate of change of $D$ on
the constraint manifold, and $D$ itself is the natural Lyapunov functional, with dissipation
function
\begin{equation}
  {\sigma_{\rm slow}^{\rm exact} = -\dot D = -\sum_e\left(g_e + \lambda h_e\right)\dot x_e
  = \sum_e \frac{\dot x_e^{\,2}}{L_e} \ge 0,}
  \label{eq:sigma_exact}
\end{equation}

vanishes exactly only at the true fixed point $g_e/h_e = -\lambda^\star$ for every edge
simultaneously (cf.\ Eq.~\eqref{eq:setpoint}) and is free of the residual budget drift implicit in
Eq.~\eqref{eq:sigma_slow}.\\

We note that a numerical implementation of Eq.~\eqref{eq:projected_flow}
should conserve $C_b$ to machine precision at every integration step, in contrast with the
$O(10^{-6})$ empirical budget tolerance reported under quenched-$\lambda$ integration
\cite{vazquez2026}; this provides a direct numerical consistency test of the present formalism.
Since the constraint is nonlinear, a first-order integrator with $\lambda$ evaluated at the
beginning of each step leaves a drift of order $(\Delta t)^2$ per step; exact conservation is restored
by a SHAKE-like retraction, which here has a closed form: because each $C_e \propto e^{2b x_e}$,
a uniform shift $x_e \to x_e + \ln(C_0/C_b)/(2b)$ after each step restores $C_b = C_0$ exactly
(Appendix~\ref{app:retraction}).\\

Table~\ref{tab:drift} compares integration schemes on a $6\times6$ jittered lattice (60 edges, floor
radius $r_{\min} = 10^{-2}$) from the same dense initial state, integrated up to $t = 60\,\kappa^{-1}$.
Three conclusions follow. First, the quenched dynamics, even with $\lambda$ set to its optimal value
$\lambda^\star$, violates the budget by $13$--$33\%$ during the transient, independently of $\Delta t$;
it returns to $C_0$ only if the final topology coincides with the one for which $\lambda^\star$ was
computed, and for $b=1$ and $\Delta t\ge 0.02$ the dynamics selects a different tree and a residual
violation of $2.7\times10^{-3}$ persists. Second, explicit Euler integration of the Gaussian dynamics,
Eq.~\eqref{eq:projected_flow}, drifts by $O(\Delta t)$ and does not self-correct, so it converges to the
optimum of a slightly different budget. Third, machine precision is reached only with the retraction;
the retraction also works with quenched $\lambda$, so exact conservation is a property of the
integrator, whereas the Gaussian multiplier is what keeps the continuous dynamics on the constraint
manifold. The large transient violations under quenched $\lambda$ also show that the two treatments
differ most during transients, as discussed next.\\

\begin{table}[t]
  \centering\small
  \setlength{\tabcolsep}{5pt}
  \caption{Maximum relative budget violation, $\max_t|C_b/C_0-1|$, for three integration schemes, and
  final violation for the scheme with a fixed multiplier. Fixed: $\lambda=\lambda^\star$ throughout.
  Gaussian: multiplier of Eq.~\eqref{eq:lambda_t} with explicit Euler steps, without and with the
  retraction of Appendix~\ref{app:retraction}. Values below $10^{-15}$ are at machine precision.}
  \label{tab:drift}
  \begin{tabular}{cccccc}
  \hline
   & & \multicolumn{2}{c}{Fixed $\lambda^\star$} & \multicolumn{2}{c}{Gaussian $\lambda(t)$}\\
  $b$ & $\kappa\Delta t$ & max & final & Euler & with retraction\\
  \hline
0.5 & 0.05 & $3.3\times10^{-1}$ & $<10^{-15}$ & $1.4\times10^{-2}$ & $<10^{-15}$ \\
0.5 & 0.02 & $3.2\times10^{-1}$ & $<10^{-15}$ & $5.6\times10^{-3}$ & $<10^{-15}$ \\
0.5 & 0.005 & $3.1\times10^{-1}$ & $<10^{-15}$ & $1.4\times10^{-3}$ & $<10^{-15}$ \\
\hline
1.0 & 0.05 & $2.3\times10^{-1}$ & $2.7\times10^{-3}$ & $3.3\times10^{-2}$ & $<10^{-15}$ \\
1.0 & 0.02 & $2.2\times10^{-1}$ & $2.7\times10^{-3}$ & $1.2\times10^{-2}$ & $<10^{-15}$ \\
1.0 & 0.005 & $2.1\times10^{-1}$ & $<10^{-15}$ & $3.0\times10^{-3}$ & $<10^{-15}$ \\
\hline
1.5 & 0.05 & $1.5\times10^{-1}$ & $<10^{-15}$ & $5.6\times10^{-2}$ & $<10^{-15}$ \\
1.5 & 0.02 & $1.4\times10^{-1}$ & $<10^{-15}$ & $2.0\times10^{-2}$ & $<10^{-15}$ \\
1.5 & 0.005 & $1.3\times10^{-1}$ & $<10^{-15}$ & $4.8\times10^{-3}$ & $<10^{-15}$ \\
\hline
  \end{tabular}
\end{table}

The
quenched treatment of Section~\ref{sec:slow} and the exact treatment derived here correspond,
respectively, to a canonical ensemble (fixed $\lambda$, fluctuating $C_b$) and a micro-canonical
ensemble (fixed $C_0$, dynamical $\lambda(t)$) for the structural sector; the two are expected to
coincide near the fixed point and in the thermodynamic limit, but may differ measurably during rapid
growth transients, precisely where staged-growth protocols operate.

\section{Fluctuating demand: a Langevin theory and the partition function over topologies}
\label{sec:fluctuations}

\subsection{Effective temperature from fluctuating demand}

Let each sink demand have unit mean and independent fluctuations of variance $\sigma^2$, with second
moment $Q = \bar I \bar I^{\mathsf T} + \operatorname{Cov}(I)$ and expected dissipation
$\langle D \rangle = \operatorname{tr}(L^{+}Q)$, as in Ref.~\cite{vazquez2026}. Two regimes must be distinguished. If demand fluctuates on the hydraulic time scale
$\tau_{\rm hyd}$, the adiabatic elimination of Section~\ref{sec:fast} replaces $D$ by $\langle D\rangle$
in $\Phi$, so that fluctuations modify the energy landscape rather than acting as noise,
$\Phi_\sigma \eqdef \langle D\rangle + \lambda C_b$. This is the mechanism of
Refs.~\cite{corson2010,katifori2010}, in which loops are already the zero-temperature optimum of the
averaged functional. Only the component of the demand fluctuations with time scales comparable to or
longer than $\tau_{\rm remod}$ acts as structural noise. In what follows $\Phi$ is understood as
$\Phi_\sigma$, and $T_{\rm eff}$ accounts for the slow component.\\

We extend the
Rayleighian dynamics of Section~\ref{sec:slow} to a Langevin equation consistent with the
fluctuation--dissipation theorem relative to the mobility already derived in Eq.~\eqref{eq:Le}:
\begin{equation}
  \dot x_e = L_e X_e + \xi_e(t), \qquad
  \langle \xi_e(t)\,\xi_{e'}(t')\rangle = 2 L_e\, T_{\rm eff}\,\delta_{ee'}\,\delta(t-t'),
  \label{eq:langevin}
\end{equation}
where the effective structural temperature $T_{\rm eff}(\sigma, b)$ is sourced by the demand
variance through the stochastic dependence of $z_e$ on $Q$. At $\sigma = 0$, $T_{\rm eff}=0$ and
Eq.~\eqref{eq:langevin} reduces to the deterministic descent of Section~\ref{sec:slow}; for
$\sigma>0$, thermal noise competes with the energetic drive toward the minimum of $\Phi_\sigma$, which for $\sigma>0$ need not be a tree.
We emphasize that Eq.~\eqref{eq:langevin} neglects multiplicative-noise (spurious-drift) corrections
arising from the $x$-dependence of $L_e$; these are subleading provided $T_{\rm eff}$ is small
compared with the local curvature scale of $\Phi$, and we flag this as a controlled approximation
rather than an exact statement. Alternatively, adding the It\^o drift $T_{\rm eff}\,\partial L_e/\partial x_e$
to Eq.~\eqref{eq:langevin} restores detailed balance with respect to $\exp(-\Phi/T_{\rm eff})$ exactly,
since the mobility matrix is diagonal. With the coupling of Section~\ref{sec:coupling}, $L_e\delta_{ee'}$
is replaced by $L^{\rm eff}_{ee'}$ in both the drift and the noise covariance, $\Phi$ by $\tilde\Phi$
(so that $T_{\rm eff}$ has units of energy), and the It\^o drift by $T_{\rm eff}\sum_{e'}\partial
L^{\rm eff}_{ee'}/\partial x_{e'}$.

\subsection{Second law and fluctuation theorem for the structural sector}
\label{sec:secondlaw}
The Langevin extension provides the structural sector with a second law in the strict sense of
stochastic thermodynamics \cite{seifert2005,seifert2012}. The probability density $p(x,t)$ of the
structural state obeys the Fokker--Planck equation
\begin{equation}
  \partial_t p = -
abla\cdot J, \qquad
  J = -L^{\rm eff}\big(p\,
abla\tilde\Phi + T_{\rm eff}
abla p\big).
  \label{eq:FP}
\end{equation}
Defining the structural free energy
$\mathcal F[p] = \int p\,\tilde\Phi\,\dd x + T_{\rm eff}\int p\ln p\,\dd x = \langle\tilde\Phi\rangle -
T_{\rm eff}S[p]$, one finds
\begin{equation}
  \dot S_{\rm tot} \eqdef \dot S[p] - \frac{1}{T_{\rm eff}}\frac{\dd\langle\tilde\Phi\rangle}{\dd t}
  = -\frac{1}{T_{\rm eff}}\frac{\dd\mathcal F}{\dd t}
  = \frac{1}{T_{\rm eff}}\int \frac{J^{\mathsf T}(L^{\rm eff})^{-1}J}{p}\,\dd x \;\ge 0,
  \label{eq:secondlaw}
\end{equation}
which is a second law for the structural sector: the total entropy production is the sum of the change
of configurational entropy and of the entropy delivered to the medium by the descent of
$\tilde\Phi$, and it is non-negative because $L^{\rm eff}$ is positive definite. Along single
trajectories the total entropy production $\Delta s_{\rm tot}$ satisfies the integral fluctuation
theorem $\langle e^{-\Delta s_{\rm tot}}\rangle = 1$ \cite{seifert2005}, of which
Eq.~\eqref{eq:secondlaw} is the average consequence. In the limit $T_{\rm eff}\to 0$ the density
concentrates on the deterministic trajectory, $\mathcal F\to\tilde\Phi$, and
Eq.~\eqref{eq:secondlaw} reduces to $T_{\rm eff}\dot S_{\rm tot}\to X^{\mathsf T}L^{\rm eff}X$, i.e.\ to
Eq.~\eqref{eq:sigma_slow} generalized by the coupling, Eq.~\eqref{eq:balance_coupled}. The stationary
state, $J=0$, is $p\propto\exp(-\tilde\Phi/T_{\rm eff})$, with detailed balance, because $L^{\rm eff}$ is
symmetric and $X$ is a gradient; for the micro-canonical treatment the same construction holds with
$L^{\rm eff}$ projected onto the manifold $C_b = C_0$, $L^{\rm eff}_\parallel = L^{\rm eff} - L^{\rm eff}hh^{\mathsf T}L^{\rm eff}/(h^{\mathsf T}L^{\rm eff}h)$, which is
symmetric and positive semi-definite with null direction normal to the manifold. The structure is that
of a dissipative (metric) GENERIC system \cite{grmela1997}.\\

Two remarks delimit this result. First, it is the second law of the effective description: $T_{\rm
eff}$ measures structural fluctuations, not the temperature of the fluid, and $\tilde\Phi$ is a
potential sustained by metabolism rather than an equilibrium free energy, so the full physical entropy
production also contains the metabolic dissipation that sustains the drive, which bounds
Eq.~\eqref{eq:secondlaw} from above. Second, the result provides a sharp test of the reciprocal
structure: if long-range mechanotransductive couplings (Section~\ref{sec:reciprocity}) were non-reciprocal,
$L_{ee'}
eq L_{e'e}$, or non-gradient, detailed balance would fail, the stationary current $J$ would
not vanish, and a housekeeping entropy production \cite{hatano2001} would appear. Time-reversal
asymmetry of radius fluctuations in pairs of vessels would then be a direct signature of active
signaling beyond hydraulics.\\

\begin{figure}[t]
  \centering
  \figfile{\textwidth}{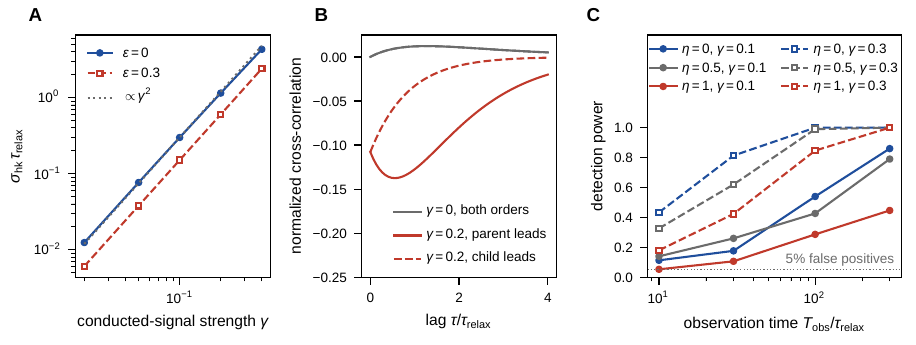}
  \caption{Time-reversal test of reciprocity, for the linearized dynamics around the optimal tree of
  the $6\times6$ lattice ($b=1$, conduction range $\xi_c = 3$ edges). (A) Housekeeping entropy
  production, multiplied by the slowest relaxation time, against the strength $\gamma$ of the
  conducted signal, Eq.~\eqref{eq:conducted}, without ($\varepsilon=0$) and with ($\varepsilon=0.3$)
  hydraulic coupling; the dotted line shows the scaling $\gamma^2$. For $\gamma=0$ it vanishes to
  machine precision in both cases. (B) Lagged cross-correlations of a parent--child pair in both time
  orders. With hydraulic coupling alone ($\gamma=0$, $\varepsilon=0.3$; grey) the two orders coincide;
  with conduction ($\gamma=0.2$, $\varepsilon=0$; red) they differ. (C) Probability of detecting $\gamma>0$ at a $5\%$ false-positive rate against observation
  time, for two signal strengths $\gamma$ and three levels of measurement noise $\eta$, given in units
  of the intrinsic fluctuation amplitude; the dotted line marks the $5\%$ false-positive level.}
  \label{fig:nonreciprocal}
\end{figure}

Figure~\ref{fig:nonreciprocal} quantifies this test for a minimal model of upstream conduction in the
spirit of Refs.~\cite{pries1998,pries2001}: each vessel also responds to the deviations from the
set-point of the vessels it feeds,
\begin{equation}
  \dot x = (\mathbb{I} + \gamma G)\,L^{\rm eff}X + \xi, \qquad
  G_{ee'} = \begin{cases} e^{-d_{ee'}/\xi_c} & e' \text{ downstream of } e,\\ 0 & \text{otherwise,}\end{cases}
  \label{eq:conducted}
\end{equation}
where $d_{ee'}$ is the distance along the tree and the noise $\xi$ obeys the fluctuation--dissipation
relation with respect to the passive mobility $L^{\rm eff}$ only. Because $G$ is not symmetric, the
mobility $(\mathbb{I}+\gamma G)L^{\rm eff}$ is not either. Linearizing around the optimal tree gives an
Ornstein--Uhlenbeck process whose stationary entropy production is entirely housekeeping
(Appendix~\ref{app:ou}). It vanishes to machine precision for $\gamma = 0$, with or without
hydraulic coupling, and grows as $\gamma^2$ (fitted exponent $1.95$; Figure~\ref{fig:nonreciprocal}A). The cross-correlations of
a parent--child pair become asymmetric under time reversal (Figure~\ref{fig:nonreciprocal}B), whereas the hydraulic coupling
alone generates correlations that are symmetric. A statistic that sums the asymmetries of all
parent--child pairs, weighted by their predicted sign, detects $\gamma = 0.3$ with power above $0.8$
after $30$ relaxation times without measurement noise, and after $100$ relaxation times when the
measurement noise equals the intrinsic fluctuation amplitude; $\gamma = 0.1$ requires about $300$
relaxation times (Figure~\ref{fig:nonreciprocal}C). In mature vasculature, where structural relaxation takes days, the test
therefore calls for imaging campaigns of months; in \emph{Physarum}, where veins respond to shear
within minutes \cite{marbach2023}, hours of imaging suffice.

\subsection{Partition function over discrete supports}

Let $S \subseteq E$ denote a connected spanning sub-network (support) of the underlying spatial graph, so that every
sink is supplied and $\beta(S)$ below is well defined, and
define the coarse-grained free energy of a support as the optimal dissipation attained on it,
\begin{equation}
  \Phi^\star(S) \eqdef \min_{\{x_e\}_{e \in S}} \Phi(x)\Big|_{C_b = C_0},
  \label{eq:Phistar}
\end{equation}
which coincides with the numerical benchmark $\widetilde D_{\rm best}(S)$ of Ref.~\cite{vazquez2026}
up to the fixed rescaling by $C_0$ and the additive constant $\lambda C_0$. Because $\alpha<1$ throughout $0<b<2$, the cost is concave in $w_e$
and, at $\sigma=0$, the constrained minimum on a support containing loops is generically not
interior: it coincides with that of one of its spanning subtrees, with some radii driven to zero.
Eq.~\eqref{eq:Phistar} must therefore be understood with a minimal radius, $x_e \ge \ln r_{\min}$,
supports whose optimum degenerates to a subtree must be identified with it to avoid double counting,
and interior minima on loopy supports exist only for $\Phi_\sigma$ with $\sigma>0$. For a Langevin process of the form~\eqref{eq:langevin} with
mobility approximately constant over the relevant fluctuations, the stationary distribution over
configurations satisfies detailed balance with respect to $P_{\rm stat}(x) \propto
\exp[-\Phi(x)/T_{\rm eff}]$.

Coarse-graining over the continuous radii within each support by a
Laplace (saddle-point) approximation yields an effective partition function over topologies,
\begin{equation}
  Z(T_{\rm eff}) = \sum_{S} \Omega(S)\, \exp\!\left[-\Phi^\star(S)/T_{\rm eff}\right], \qquad
  \Omega(S) \propto {(2\pi T_{\rm eff})^{d_S/2}}\det\!\left[\frac{\partial^2 \Phi}{\partial x^2}\right]^{-1/2}_{S},
  \label{eq:Z}
\end{equation}
where $d_S$ is the number of continuous degrees of freedom on $S$. The ensemble must be chosen
consistently: in the micro-canonical form, Eq.~\eqref{eq:Phistar} is minimized on $C_b = C_0$, the
Hessian is restricted to the tangent space of the constraint and $d_S = |S|-1$; in the canonical form,
$\lambda$ is fixed, $\Phi^\star(S)$ is the unconstrained minimum of $D+\lambda C_b$ on $S$ (whose budget
then differs from support to support) and $d_S = |S|$. Since $|S| = \beta(S) + n - 1$, the prefactor
contains $T_{\rm eff}^{\beta(S)/2}$: an explicitly temperature-dependent entropic weight that favours
loops as $T_{\rm eff}$ grows and must be kept in Eq.~\eqref{eq:beta_avg}. The entropic prefactor
$\Omega(S)$ is the network analogue of a vibrational entropy, obtained from the Gaussian fluctuations
of $\Phi$ around its minimum on $S$ (Appendix~\ref{app:laplace}). The Laplace approximation requires the Hessian restricted to $S$ to be positive definite and
$T_{\rm eff}$ to be small compared with its smallest eigenvalue.

\subsection{The loop count as a thermal order parameter}
\label{sec:loops}

\begin{figure}[t]
  \centering
  \figfile{\textwidth}{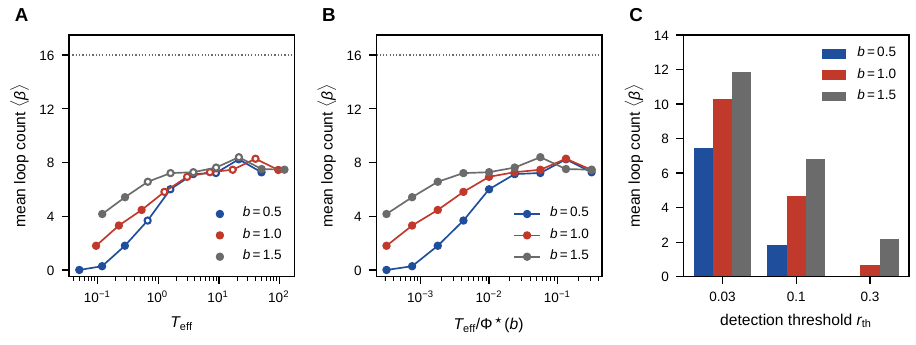}
  \caption{Test of the data-collapse hypothesis by Metropolis sampling on a $5\times5$ jittered
  lattice; the dotted line marks the maximum number of independent loops (16). (A) Mean loop count
  against $T_{\rm eff}$ for three cost exponents. Filled symbols mark temperatures at which the
  equipartition calibration agrees with $T_{\rm eff}$ within $35\%$. (B) The same data against
  $T_{\rm eff}/\Phi^\star(b)$. (C) Mean loop count at $T_{\rm eff}/\Phi^\star = 1.8\times10^{-3}$ for
  three detection thresholds $r_{\rm th}$.}
  \label{fig:collapse}
\end{figure}

With $\beta(S) = |S| - n + 1$ the number of independent loops of a support, its thermal average is
\begin{equation}
  \langle \beta \rangle (T_{\rm eff}) = \frac{1}{Z}\sum_S \beta(S)\, \Omega(S)\,
  \exp\!\left[-\Phi^\star(S)/T_{\rm eff}\right].
  \label{eq:beta_avg}
\end{equation}
As $T_{\rm eff}\to 0$, $Z$ is dominated by the lowest-$\Phi^\star$ support, which is generically a
tree for concave transport costs \cite{banavar2000,bohn2007,durand2007}, recovering
$\langle\beta\rangle \to 0$; as $T_{\rm eff}\to\infty$, all supports are weighted by $\Omega(S)$
alone, which now grows as $T_{\rm eff}^{\beta(S)/2}$, and since loopy connected subgraphs vastly outnumber spanning trees combinatorially,
$\langle\beta\rangle$ saturates at a $b$-independent value, which in the tests of Figure~\ref{fig:collapse} is set by
the detection threshold and lies at about half the maximum loop count; this limit lies
outside the validity of the Laplace approximation and should be read as a qualitative extrapolation, to
be checked by direct enumeration on small graphs. This
reinterprets the previously reported monotonic growth of surviving loops with demand fluctuation
\cite{vazquez2026} as an order--disorder crossover, which can sharpen into a genuine transition only in the limit of large graphs \cite{kaiser2020}, with $\langle\beta\rangle$ playing the role of
the order parameter and $T_{\rm eff}$ that of the control parameter.\\

This picture suggests a specific test that goes beyond fitting $\beta$ against $\sigma$ for each cost
exponent $b$: if $T_{\rm eff}(\sigma,b)$ is calibrated independently from the
stationary variance of structural fluctuations in simulation, $T_{\rm eff} = \langle(\delta x_e)^2\rangle / [H_S^{-1}]_{ee}$, with
$H_S = \partial^2\Phi/\partial x\,\partial x$ the Hessian at the minimum on $S$ (equipartition; for a single
decoupled edge, $T_{\rm eff} = k_e\langle(\delta x_e)^2\rangle$ with $k_e = \partial^2\Phi/\partial x_e^2$) at a reference edge,
 then $\langle\beta\rangle$ plotted against this
calibrated $T_{\rm eff}$ should collapse the family of curves obtained for different values of $b$
onto a single master curve, to the extent that the $b$-dependence of the
free-energy differences $\Phi^\star(S)-\Phi^\star(S')$ and of $\Omega(S)$ is captured by the single
scale $T_{\rm eff}$; systematic deviations from collapse would then measure the residual
$b$-dependence of the landscape itself.
This hypothesis can be tested on existing simulation data, provided the stationary radius
fluctuations were recorded.\\

We carried out this test by Metropolis sampling of $P\propto\exp[-(D+\lambda^\star C_b)/T_{\rm eff}]$ on a
$5\times5$ jittered lattice (40 edges, at most 16 independent loops) for $b = 0.5$, $1$ and $1.5$,
counting as active the vessels with $r_e > r_{\rm th} = 0.1$ ($r_{\min} = 10^{-2}$). The results
(Figure~\ref{fig:collapse}) do not support the collapse, in contrast with the curve collapse reported for
fluctuation-driven transitions in coupled Hu--Cai networks \cite{kramer2020}, where the control
parameter is the fluctuation amplitude itself rather than a calibrated temperature. (i) The equipartition calibration is valid only in
the harmonic regime, up to $T_{\rm eff}/\Phi^\star \approx 2\times10^{-3}$; beyond it the reference edge
leaves its well and the calibrated temperature diverges. (ii) In the crossover region the curves do not
collapse, neither against $T_{\rm eff}$ nor against $T_{\rm eff}/\Phi^\star$: at
$T_{\rm eff}/\Phi^\star = 3\times10^{-4}$ one finds $\langle\beta\rangle = 0$, $1.8$ and $4.2$ for
$b = 0.5$, $1$ and $1.5$. The $b$-dependence is dominated by the cost of small vessels, $C_e\propto
r_e^{2b}$, which makes the thermal activation of loops cheaper at larger $b$ and is not absorbed by a
single temperature scale. (iii) At high $T_{\rm eff}$ the curves merge on a $b$-independent plateau,
$\langle\beta\rangle\approx7.5$, whose value, like $\langle\beta\rangle$ everywhere, depends strongly on
$r_{\rm th}$. A meaningful order parameter therefore requires either a threshold-free definition, for
instance a conductance-weighted measure of redundancy, or an explicit threshold matched to the
resolution of the imaging. These tests use purely thermal noise ($\sigma = 0$); demand fluctuations
enter through $\Phi_\sigma$ and must be tested separately.

\section{Discussion and conclusion}
\label{sec:discussion}

The classical minimum-entropy-production theorem guarantees a unique stationary state only near
equilibrium, in the strictly linear regime, and on a fixed support. Both conditions hold for the
fast hydraulic sector (Section~\ref{sec:fast} and Eq.~\eqref{eq:sigma_fast}), where the physical flow field is exactly the unique
minimizer of dissipation on a fixed conductance network. Neither condition holds for the slow
structural sector once pruning and growth are allowed to change the admissible support: $\Phi$ is
strictly convex in $\{x_e\}$ only on a fixed tree, and the map from initial to final support is not
unique. This is the thermodynamic origin of the previously reported topological degeneracy and
developmental-history dependence \cite{vazquez2026}; it is an expected consequence of leaving the
domain of validity of the classical uniqueness theorem once the constitutive support itself is
allowed to vary, rather than a numerical artifact of the search procedure.\\

In summary, we have shown that: (i) the hydraulic sector is an exact linear Onsager system with
automatic reciprocity and an associated minimum-dissipation theorem; (ii) the structural adaptation
rule and its Lyapunov inequality admit a gradient-flow representation through a Rayleighian variational principle, which
identifies the structural Onsager coefficient $L_e = \kappa/[2\lambda b\, {C_e}(z_e+1)]$ in closed
form and predicts an inverse scaling of structural mobility with local maintenance cost; (iii) exact
enforcement of the metabolic budget constraint requires a Gaussian-constraint correction to the
multiplier, Eq.~\eqref{eq:lambda_t}, which takes an explicit closed form, distinguishing a canonical from a micro-canonical treatment of
the structural dynamics; (iv) fluctuating demand admits a Langevin extension whose stationary
measure defines a partition function over discrete network topologies, reframing loop selection as
an order--disorder crossover, although the data-collapse hypothesis suggested by this picture is
not supported by the numerical test of Section~\ref{sec:loops}; and (v) the fluid displaced by wall motion couples both sectors through a symmetric,
non-diagonal effective mobility, and the Langevin extension satisfies a second law and an integral
fluctuation theorem for the structural sector. Together these results
recast a variational optimization result as a predictive non-equilibrium thermodynamic theory, with
falsifiable consequences, namely the mobility--cost scaling, the exact-conservation numerical test, the reciprocal, negative cross-mobilities of hydraulically connected vessels, and the
time-reversal symmetry of their fluctuations, that go beyond what is accessible from the original
formulation.\\

Natural extensions include relaxing the diagonal (purely local) structural Onsager matrix to admit
long-range mechanotransductive couplings $L_{ee'}$ beyond the hydraulic coupling of
Section~\ref{sec:coupling}, testing Onsager reciprocity for such couplings
directly against vascular data, and carrying out the small-network enumeration outlined in
Section~\ref{sec:fluctuations} to validate the partition-function description quantitatively, with a
threshold-free order parameter for loop selection.

\appendix
\setcounter{table}{0}
\renewcommand{\thetable}{A.\arabic{table}}
\renewcommand{\theHtable}{A.\arabic{table}}

\section{Derivations}
\label{app:derivations}

\subsection{Notation}
\label{app:notation}

Table~\ref{tab:notation} lists the symbols used in the main text.
\begin{table}[h]
  \centering\small
  \caption{Notation (reduced units).}
  \label{tab:notation}
  \begin{tabular}{ll}
  \hline
  $r_e$, $\ell_e$, $x_e=\ln r_e$ & radius, length, logarithmic radius of edge $e$\\
  $w_e = r_e^4/\ell_e$, $f_e$, $\Delta p_e$ & conductance, flow, pressure drop ($f_e = w_e\Delta p_e$)\\
  $\tau_e = f_e/r_e^3$, $\tau_{\rm set}$, $z_e$ & wall shear stress, set-point, $z_e = |\tau_e|/\tau_{\rm set}$\\
  $b$, $\alpha = b/2$ & cost exponent\\
  $c_e = \ell_e^{3b/2}$, $C_e = c_e w_e^\alpha = r_e^{2b}\ell_e^b$ & length factor, maintenance cost of edge $e$\\
  $C_b = \sum_e C_e$, $C_0$, $\lambda$ & total cost, budget, metabolic multiplier\\
  $D$, $\Phi = D+\lambda C_b$ & dissipation, composite functional\\
  $\kappa$, $L_e$, $X_e = -\partial\Phi/\partial x_e$ & rate constant, structural mobility, force\\
  $\tau_\Phi$, $\tilde\Phi = \tau_\Phi\Phi$, $\Gamma_e$ & metabolic time, structural potential, bare mobility\\
  $v_e = \nu r_e^2\ell_e$, $K$, $L^{\rm eff}$ & lumen volume, hydraulic friction, effective mobility\\
  $\varepsilon_e = \Gamma_e K_{ee}$ & coupling strength\\
  $T_{\rm eff}$, $\gamma$, $\xi_c$ & structural temperature, conducted-signal strength and range\\
  \hline
  \end{tabular}
\end{table}

\subsection{Shear set-point}
\label{app:setpoint}

This subsection derives Eq.~\eqref{eq:setpoint}. With the Lagrangian $D + \lambda(C_b - C_0)$ and
$\partial D/\partial w_e = -(\Delta p_e)^2$ (Appendix~\ref{app:envelope}), stationarity gives
$(\Delta p_e)^2 = \lambda\alpha\, c_e w_e^{\alpha-1} = \lambda\alpha\, C_e/w_e$. Since
$\tau_e = w_e\Delta p_e/r_e^3$,
\[
  \tau_e^2 = \frac{w_e^2(\Delta p_e)^2}{r_e^6} = \frac{\lambda\alpha\,C_e w_e}{r_e^6}
  = \lambda\alpha\, r_e^{2b-2}\ell_e^{\,b-1},
\]
which is Eq.~\eqref{eq:setpoint}. For an arbitrary state the same algebra defines $z_e$ through
$(\Delta p_e)^2 = z_e^2\lambda\alpha\, C_e/w_e$.

\subsection{Envelope theorem}
\label{app:envelope}

At fixed demands the flow minimizes $\sum_e f_e^2/w_e$ subject to
Kirchhoff's law. The derivative of the minimum with respect to $w_e$ equals the partial derivative of
the Lagrangian at the optimum, $\partial D/\partial w_e = -f_e^2/w_e^2 = -(\Delta p_e)^2$; the implicit
dependence of the flows on $w_e$ does not contribute.

\subsection{Structural force and Lyapunov identity}
\label{app:force}

This subsection derives Eqs.~\eqref{eq:Phidot_reported}, \eqref{eq:dPhidx} and \eqref{eq:sigma_slow}.
With $\dd w_e/\dd x_e = 4w_e$ and the results of Appendices~\ref{app:setpoint} and~\ref{app:envelope},
\[
  \frac{\partial\Phi}{\partial x_e} = 4w_e\Big[-(\Delta p_e)^2 + \lambda\alpha\, c_e w_e^{\alpha-1}\Big]
  = 4\lambda\alpha\, C_e(1-z_e^2).
\]
With $4\alpha = 2b$ and the rule~\eqref{eq:rule},
$\dot\Phi = \sum_e(\partial\Phi/\partial x_e)\,\kappa(z_e-1) = -2b\lambda\kappa\sum_e C_e(z_e-1)^2(z_e+1)$.

\subsection{Gaussian multiplier}
\label{app:gauss}

This subsection derives Eqs.~\eqref{eq:lambda_t}--\eqref{eq:sigma_exact}. From
Appendix~\ref{app:setpoint},
$z_e^2 = w_e(\Delta p_e)^2/(\lambda\alpha C_e) = s_e^2/\lambda$, with $s_e$ independent of $\lambda$.
Since $h_e = \partial C_b/\partial x_e = 2bC_e$, the constraint reads
$\sum_e h_e\dot x_e = 2b\kappa\sum_e C_e(s_e/\sqrt\lambda - 1) = 0$, whose solution is
Eq.~\eqref{eq:lambda_t}. Along the constrained flow,
$\dot D = \sum_e g_e\dot x_e = \sum_e(g_e+\lambda h_e)\dot x_e = -\sum_e \dot x_e^2/L_e$, because
$\dot x_e = -L_e(g_e+\lambda h_e)$ and $\sum_e h_e\dot x_e = 0$.

\subsection{Retraction}
\label{app:retraction}

Since $C_e \propto e^{2bx_e}$, a uniform shift $\delta$ of the free edges
multiplies their cost by $e^{2b\delta}$. With $C_{\rm f}$ and $C_{\rm c}$ the costs of free and clamped
(floor) edges, $\delta = \ln[(C_0 - C_{\rm c})/C_{\rm f}]/(2b)$ restores $C_b = C_0$ exactly.

\subsection{Coupled sectors}
\label{app:coupled}

This subsection derives Eqs.~\eqref{eq:wall}--\eqref{eq:balance_coupled}. Stationarity of
Eq.~\eqref{eq:R_joint} with respect to $f_e$ gives $f = WB^{\mathsf T}\phi$ ($W = {\rm diag}\,w$);
inserted in the constraint, $L\phi = I - AV'\dot x$. The static part, $\phi^{(0)} = L^+I$, is the
pressure of Section~\ref{sec:model}; the dynamic part, with the source pressure fixed, is
$\phi^{\rm dyn} = -L_{\rm r}^{-1}A_{\rm r}V'\dot x$. Stationarity with respect to $\dot x_e$ gives
Eq.~\eqref{eq:wall}; keeping the dynamic pressure only, $\dot x = \Gamma(X - K\dot x)$, i.e.\
Eq.~\eqref{eq:coupled}. The flows induced by wall motion, $f^{\rm dyn} = WB_{\rm r}^{\mathsf T}\phi^{\rm dyn}$,
dissipate
$\sum_e (f_e^{\rm dyn})^2/w_e = \phi^{{\rm dyn}\,\mathsf T}L_{\rm r}\phi^{\rm dyn} = \dot x^{\mathsf T}K\dot x$,
which gives Eq.~\eqref{eq:balance_coupled}.

\subsection{Sign of the cross-mobilities}
\label{app:sign}

Writing
$L^{\rm eff} = \Gamma^{1/2}(\mathbb{I} + \Gamma^{1/2}K\Gamma^{1/2})^{-1}\Gamma^{1/2} = \Gamma - \Gamma K\Gamma +
O(K^2)$, the off-diagonal elements are $-\Gamma_eK_{ee'}\Gamma_{e'}$ to first order. $L_{\rm r}$ is the
Laplacian with the source grounded; on each connected component of the graph without the source it is
an irreducible M-matrix, whose inverse has strictly positive entries, and it is block-diagonal across
components. Hence $K_{ee'} = v'_ev'_{e'}\sum_{ij}A_{ie}(L_{\rm r}^{-1})_{ij}A_{je'} > 0$ for $e$, $e'$ in the
same branch, and $K_{ee'} = 0$ otherwise.

\subsection{Second law}
\label{app:secondlaw}

This subsection derives Eq.~\eqref{eq:secondlaw}. With Eq.~\eqref{eq:FP},
\[
  \frac{\dd\mathcal F}{\dd t} = \int \partial_t p\,\big(\tilde\Phi + T_{\rm eff}\ln p\big)\,\dd x
  = \int J^{\mathsf T}
abla\big(\tilde\Phi + T_{\rm eff}\ln p\big)\,\dd x
  = -\int\frac{J^{\mathsf T}(L^{\rm eff})^{-1}J}{p}\,\dd x \le 0,
\]
where the second step integrates by parts (the term $\int\partial_t p\,\dd x$ vanishes) and the third
uses $J = -p\,L^{\rm eff}
abla(\tilde\Phi + T_{\rm eff}\ln p)$.

\subsection{Laplace prefactor}
\label{app:laplace}

This subsection derives the prefactor of Eq.~\eqref{eq:Z}. Expanding
$\Phi = \Phi^\star(S) + \tfrac12\delta x^{\mathsf T}H_S\delta x$ around the minimum on $S$,
$\int e^{-\Phi/T_{\rm eff}}\dd^{d_S}x \simeq e^{-\Phi^\star(S)/T_{\rm eff}}(2\pi T_{\rm eff})^{d_S/2}(\det H_S)^{-1/2}$.

\subsection{Ornstein--Uhlenbeck entropy production}
\label{app:ou}

For $\dd\,\delta x = -\mathsf A\,\delta x\,\dd t + \sqrt{2}\,\mathsf B\,\dd W$ with
$\mathsf N = \mathsf B\mathsf B^{\mathsf T}$, the stationary covariance solves
$\mathsf A\mathsf C + \mathsf C\mathsf A^{\mathsf T} = 2\mathsf N$, and the stationary current is
$J = p\,v$ with $v = -(\mathsf A - \mathsf N\mathsf C^{-1})\delta x$. The stationary entropy production,
which is entirely housekeeping, is
\[
  \sigma_{\rm hk} = \langle v^{\mathsf T}\mathsf N^{-1}v\rangle
  = \operatorname{tr}\!\big[(\mathsf A - \mathsf N\mathsf C^{-1})^{\mathsf T}\mathsf N^{-1}
  (\mathsf A - \mathsf N\mathsf C^{-1})\,\mathsf C\big].
\]
For Eq.~\eqref{eq:conducted}, $\mathsf A = (\mathbb{I}+\gamma G)L^{\rm eff}H$ and
$\mathsf N = T_{\rm eff}L^{\rm eff}$, with $H$ the Hessian of $\tilde\Phi$. At $\gamma = 0$,
$\mathsf C = T_{\rm eff}H^{-1}$ solves the Lyapunov equation and $\mathsf A = \mathsf N\mathsf C^{-1}$, so
that $v\equiv 0$ and $\sigma_{\rm hk} = 0$: detailed balance holds for any symmetric $L^{\rm eff}$. The
lagged correlations are $\langle\delta x(t+\tau)\,\delta x(t)^{\mathsf T}\rangle = e^{-\mathsf A\tau}\mathsf C$;
their antisymmetric part vanishes identically under detailed balance.

\section{Numerical methods}
\label{app:numerics}

\subsection{Networks}

Square lattices of $m\times m$ nodes ($m = 6$, 60 edges; $m = 5$, 40 edges) with
node positions perturbed uniformly by up to $0.2$ lattice units, so that lengths vary; the source is
a corner node, every other node withdraws one unit of flow, and pressures are obtained from the
grounded Laplacian. A floor radius $r_{\min} = 10^{-2}$ prevents singular conductances; edges at the
floor with negative velocity are clamped.

\subsection{Deterministic dynamics}

This subsection describes the runs of Table~\ref{tab:drift}. The rule~\eqref{eq:rule} with $\kappa = 1$ is
integrated by explicit Euler up to $t = 60$ from a dense initial state, $x_e(0) \sim \mathcal N(0,0.1^2)$.
The multiplier is either fixed at the value $\lambda^\star$ of a reference run
($\Delta t = 0.005$, Gaussian multiplier with retraction) or given by Eq.~\eqref{eq:lambda_t}; the
retraction is that of Appendix~\ref{app:retraction}.

\subsection{Langevin simulations}

For Figure~\ref{fig:mobility} we use Euler--Maruyama integration around the
optimal tree with $\Delta t = 2\times10^{-3}$, $4\times10^4$ steps, $T_{\rm eff} = 10^{-4}\lambda^\star\bar C$
(with $\bar C$ the mean edge cost) and fixed $\lambda^\star$; mobilities are estimated from the covariance
of one-step increments, $\hat L = \operatorname{Cov}(\Delta x)/(2T_{\rm eff}\Delta t)$. The coupling
strength is set by rescaling $
u$ so that the median of $\Gamma_eK_{ee}$ equals $\varepsilon$.

\subsection{Monte Carlo sampling}

For Figure~\ref{fig:collapse} we use Metropolis sampling of
$\exp[-(D + \lambda^\star C_b)/T_{\rm eff}]$ with single-edge Gaussian proposals of width $0.35$ in $x_e$,
reflected at the floor; $D$ is updated with the Sherman--Morrison formula and recomputed every 50
sweeps; 1500 sweeps per temperature, the first $30\%$ discarded. Loops are counted as the cycle rank of
the subgraph of edges with $r_e > r_{\rm th}$.

\subsection{Reciprocity test}

For Figure~\ref{fig:nonreciprocal}, the linearized dynamics of Appendix~\ref{app:ou} is
simulated exactly in discrete time, $\delta x_{k+1} = e^{-\mathsf A\Delta}\delta x_k + \zeta_k$ with
$\operatorname{Cov}\zeta = \mathsf C - e^{-\mathsf A\Delta}\mathsf C e^{-\mathsf A^{\mathsf T}\Delta}$, sampled at
$\Delta = \tau_{\rm relax}/2$ with $\tau_{\rm relax}^{-1}$ the smallest real part of the eigenvalues of
$\mathsf A$. Measurement noise of standard deviation $\eta\,\mathsf C_{ee}^{1/2}$ is added independently to
each sample. The test statistic is $\sum_{(e,e')}{\rm sign}(a_{ee'})\,\hat a_{ee'}$ over parent--child
pairs, with $\hat a_{ee'}$ the normalized antisymmetric part of the lagged correlation at
$\tau = \tau_{\rm relax}$ and $a_{ee'}$ its theoretical value; the threshold is the $95\%$ quantile of the
statistic for $\gamma = 0$, and the power is estimated from 300 independent realizations per point.

\subsection{Code availability}

All calculations use Python (NumPy, SciPy); the scripts reproduce every figure and
table, and are available at \texttt{https://github.com/Danpc11/setpoint-thermo}.

\section*{Acknowledgements}
  This work was supported by DGAPA-PAPIIT under Grant No.\ IN234029 and by SECIHTI under Grant
  No.\ CBF-2025-G-789.

\end{document}